\documentclass[journal=aamick,manuscript=article]{achemso}

\usepackage{color}
\usepackage{graphicx}
\usepackage{dcolumn}
\usepackage{bm}
\usepackage{amsmath}
\usepackage{wrapfig}
\usepackage{graphicx}
\usepackage{xr}
\setkeys{acs}{maxauthors=0}
\setkeys{acs}{chaptertitle=true}
\setkeys{acs}{articletitle=true}

\usepackage{xr}
\makeatletter
\newcommand*{\addFileDependency}[1]{
  \typeout{(#1)}
  \@addtofilelist{#1}
  \IfFileExists{#1}{}{\typeout{No file #1.}}
}
\makeatother

\newcommand*{\myexternaldocument}[1]{%
    \externaldocument{#1}%
    \addFileDependency{#1.tex}%
    \addFileDependency{#1.aux}%
}

\myexternaldocument{supplemental_material}

\title{Quantifying Mn diffusion through transferred versus directly-grown graphene barriers
}

\author{Patrick J. Strohbeen}
\affiliation{Materials Science and Engineering, University of Wisconsin Madison, Madison, Wisconsin, 53706, USA}

\author{Sebastian Manzo}
\affiliation{Materials Science and Engineering, University of Wisconsin Madison, Madison, Wisconsin, 53706, USA}

\author{Vivek Saraswat}
\affiliation{Materials Science and Engineering, University of Wisconsin Madison, Madison, Wisconsin, 53706, USA}

\author{Katherine Su}
\affiliation{Materials Science and Engineering, University of Wisconsin Madison, Madison, Wisconsin, 53706, USA}

\author{Michael S. Arnold}
\affiliation{Materials Science and Engineering, University of Wisconsin Madison, Madison, Wisconsin, 53706, USA}

\author{Jason K. Kawasaki}
\email{jkawasaki@wisc.edu}
\affiliation{Materials Science and Engineering, University of Wisconsin Madison, Madison, Wisconsin, 53706, USA}

\date{\today}
\begin{document}
\begin{abstract}
We quantify the mechanisms for manganese (Mn) diffusion through graphene in Mn/graphene/Ge (001) and Mn/graphene/GaAs (001) heterostructures for samples prepared by graphene layer transfer versus graphene growth directly on the semiconductor substrate. These heterostructures are important for applications in spintronics; however, challenges in synthesizing graphene directly on technologically important substrates such as GaAs necessitate layer transfer and anneal steps, which introduce defects into the graphene. \textit{In-situ} photoemission spectroscopy measurements reveal that Mn diffusion through graphene grown directly on a Ge (001) substrate is 1000 times lower than Mn diffusion into samples without graphene ($D_{gr,direct} \sim 4\times10^{-18}$cm$^2$/s, $D_{no-gr} \sim 5 \times 10^{-15}$ cm$^2$/s at 500$^\circ$C). Transferred graphene on Ge suppresses the Mn in Ge diffusion by a factor of 10 compared to no graphene ($D_{gr,transfer} \sim 4\times10^{-16}cm^2/s$). For both transferred and directly-grown graphene, the low activation energy ($E_a \sim 0.1-0.5$ eV) suggests that Mn diffusion through graphene occurs primarily at graphene defects. This is further confirmed as the diffusivity prefactor, $D_0$, scales with the defect density of the graphene sheet. Similar diffusion barrier performance is found on GaAs substrates; however, it is not currently possible to grow graphene directly on GaAs. Our results highlight the importance of developing graphene growth directly on functional substrates, to avoid the damage induced by layer transfer and annealing.\\

\textbf{keywords:} graphene, diffusion, epitaxy, photoemission, defect

\end{abstract}

\maketitle

\section{Introduction}

Graphene is a promising monolayer diffusion barrier for applications such as oxidation resistance \cite{topsakal2012graphene}, metal contacts to semiconductors \cite{roy2013improving, hong2014graphene}, and semiconductor spintronics \cite{cobas2012graphene}. Compared to conventional diffusion barriers that typically need to be tens of nanometers thick, graphene can remain effective as a diffusion barrier down to thicknesses of several atomic layers \cite{roy2013improving}, and in select cases a single atomic layer \cite{hong2014graphene}. This extreme thinness is highly attractive for applications where a nearly electronically transparent interface is desired, for example, spin injection via tunneling \cite{cobas2012graphene}. Moreover, recent reports of ``remote epitaxy'' suggest it is possible to synthesize epitaxial films through a monolayer graphene barrier \cite{lu2018remote, kim2017remote, du2021epitaxy}, opening the possibility of epitaxial metal/graphene/semiconductor heterostructures that would otherwise not be stable due to significant metal/semiconductor interdiffusion \cite{sands1990stable}.

A fundamental challenge, however, is that graphene cannot be synthesized directly on arbitrary substrates. While graphene can be grown by chemical vapor deposition (CVD) on Ge and noble metal surfaces (e.g., Cu, Ag, Pt), the high synthesis temperatures and catalytic reactions at the surface currently preclude growth directly on technologically important compound semiconductor substrates such as GaAs and InP. The use of non-native substrates necessitates post-synthesis large-area graphene layer transfer and annealing, which can damage the graphene by introducing contaminants such as oxides and polymer residues, and creating tears, wrinkles, and point defects on the larger length scales relevant to device applications\cite{ma2019xferreview}. Previous studies have explored the qualitative impacts of defects on graphene diffusion barriers \cite{oh2015effects, roy2013improving, yuan2013grgb}. However, the specific effect of the layer transfer process, elevated temperatures during growth, and a quantification of the diffusivity through graphene has not yet been reported.

Here we quantify the effectiveness of layer-transferred vs directly-grown graphene as a monolayer diffusion barrier for Mn films, grown on GaAs and Ge substrates. The choice of Mn on a semiconductor is motivated by applications in spintronics. For example, the Mn-containing half metallic ferromagnets NiMnSb and Co$_2$MnSi are promising materials for spin injection into GaAs \cite{farshchi2013spin, dong2005spin}. Sample structures consisted of a thin Mn film ($1.64\times10^{16}$ atoms/cm$^2$, $\sim 24$ \AA) grown by molecular-beam epitaxy (MBE) on graphene-terminated Ge or GaAs (001) substrates. We evaluate two graphene preparations: (1) epitaxial graphene that is directly grown on a Ge substrate and (2) transferred graphene that is grown on a sacrificial Cu foil and then wet transferred to a Ge or GaAs substrate. We show that the combination of graphene transfer and pre-growth annealing creates tears and pinholes in the graphene, as quantified by Raman spectroscopy and identified by scanning electron micrsocopy (SEM) and atomic force microscopy (AFM). \textit{In-situ} x-ray photoemission spectroscopy (XPS) measurements reveal that both transferred and directly-grown graphene suppress diffusion into the semiconductor substrate. Directly-grown graphene on Ge significantly outperforms the layer transferred versions, decreasing the diffusivity at 500$^\circ$C two orders of magnitude compared to the layer transferred version. We quantify the effective diffusivities and activation energies for Mn diffusion on graphene-terminated Ge and GaAs, and discuss mechanisms for diffusion with and without tears and pinholes.

\section{Experimental}

\textbf{CVD growth of graphene on Ge (001) and Cu foils.} We use two preparations of graphene: graphene that is grown by chemical vapor deposition (CVD) directly on Ge (001), and graphene that is grown by CVD on Cu foils and then layer transferred onto Ge (001) or GaAs (001). For CVD graphene growth on Ge (001) we followed growth conditions described in Refs. \cite{kiraly2015, wang2013grge}. CVD graphene on Cu foils was performed at 1050 $^\circ$C using ultra high purity CH$_4$, as described in Ref. \cite{du2021epitaxy}.

\textbf{Graphene transfer onto Ge (001) and GaAs (001).} Our graphene transfer process is similar to other polymer-assisted wet transfers reported in previous works \cite{park2018, du2021epitaxy}. The graphene/Cu foils were cut into 5 by 5 mm pieces and flattened using cleaned glass slides. Approximately 300 nm of 495K C2 PMMA (Chlorobenzene base solvent, 2\% by wt., MicroChem) was spin coated on the graphene/Cu foil substrate at 2000 RPM for 2 minutes and left to cure at room temperature for 24 hours. Graphene on the backside of the Cu foil was removed via reactive ion etching using 90 W O$_2$ plasma at a pressure of 100 mTorr for 30s. The Cu foil was then etched by placing the PMMA/graphene/Cu foil stack on the surface of an etch solution containing 1-part ammonium persulfate (APS-100, Transene) and 3-parts H$_2$O. After 16 hours of etching at room temperature, the floating PMMA/graphene membrane was scooped up with a clean glass slide and sequentially transferred into five 5-minute water baths to rinse the etch residuals. Prior to scooping the graphene with the GaAs (001) substrate, the native oxide is etched in a 10\% HCl bath for 30 seconds. The GaAs is then rinsed in isopropanol and nitrogen dried. Prior to scooping with the Ge (001) substrate, the native oxide is etched in a hot (90$^\circ$C) RO water bath for 15 minutes. Immediately after the etching process, the substrate is used to scoop the PMMA/graphene membrane from the final water bath. To remove water at the graphene/substrate interface, samples were baked in air at 50$^\circ$C for 5 minutes, then slowly ramped to 150$^\circ$C and baked for another 10 minutes. The PMMA was dissolved by submerging the sample in an acetone bath at 80$^\circ$C for 3 hours. Samples are then rinsed in isopropanol and dried with nitrogen.

\textbf{Graphene surface preparation and manganese film growth.} Bare Ge substrates were sequentially degreased for 15 minutes in acetone, 15 minutes in isopropanol baths, and then etched in reverse-osmosis (RO) water at 90$^{\circ}$C for 30 minutes. No solvent treatments were performed to the bare GaAs substrates. All samples were indium-bonded to a molybdenum sample holder to ensure good thermal contact. Samples were outgassed in a high vacuum load lock ($p < 5\times 10^{-8}$ Torr) for at least 1 hour at 150$^{\circ}$C prior to transferring into the ultra-high vacuum (UHV) molecular beam epitaxy (MBE) chamber ($p < 5\times10^{-10}$ Torr).

In the MBE chamber, the graphene-terminated samples were further outgassed at $\sim 300^\circ$C to remove residual organic contaminants. Select graphene samples were characterized by SEM, AFM, and Raman spectroscopy after the $\sim 300^\circ$C outgas (labelled ``pre-anneal'' in Fig. \ref{graphene_char}). Following the $\sim 300^\circ$C outgas, graphene samples were annealed at roughly 600$^{\circ}$C to remove native oxides. Removal of surface oxides was confirmed via reflection high energy electron diffraction (RHEED). For the bare GaAs substrate, a thick GaAs buffer layer was grown to smoothen the surface before Mn growth. No buffer layers were grown on the other samples.

Thin Mn films ($1.64\times10^{16}$ atoms/cm$^2$) were grown at room temperature, using an elemental effusion cell. The thickness was measured using an \textit{in-situ} quartz crystal microbalance (QCM) that has been calibrated through \textit{ex-situ} Rutherford backscattering spectroscopy (RBS).

\textbf{In-situ photoemission.} Immediately following growth, samples were transferred through an ultrahigh vacuum manifold ($p< 5\times10^{-9}$ Torr) to a connected photoemission spectroscopy chamber. X-ray photoemission specroscopy (XPS) measurements were conducted using a non-monochromated Omicron DAR-400 dual anode x-ray source (Al $K \alpha$, $h\nu= 1486.6$ eV) and an Omicron EA125 hemispherical analyzer. The total energy resolution of the analyzer in the experimental configuration was measured to be 1.08 eV using an Au standard. The photoemission chamber pressure was maintained at $< 3\times10^{-10}$ Torr during the measurements.

At each anneal step the samples were ramped from room temperature to the desired anneal setpoint temperature over 5 minutes. The sample was held at temperature ($\pm10^{\circ}$C) for 5 minutes and then rapidly cooled to room temperature. Sample temperature was measured by a thermocouple attached to the sample manipulator, roughly 1 cm away from the sample. The thermocouple was calibrated using the melting point of indium and the arsenic desorption temperature of As-capped GaAs. The sample was transferred immediately following the cool down back into the photoemission stage before beginning the next XPS measurement.

\textbf{Raman and scanning electron microscopy.}
Graphene was characterized using field emission scanning electron microscopy (SEM) (Zeiss LEO 1530 Gemini). Raman spectroscopy was performed using a 532 nm wavelength laser (Horiba LabRAM HR Evolution Spectrometer). The laser power is fixed at 10 mW and a 300 gr/mm diffraction grating is used for every scan. This results in a spectral resolution of 2.5 cm$^{-1}$ and a spatial resolution of 0.5 $\mu$m. Each scan presented is an average spectrum consisting of 400 scans over a representative 100 $\mu$m$^{2}$ area of the graphene surface.

\section{Results and Discussion}

\begin{figure}[ht]
    \includegraphics[width=.95\linewidth]{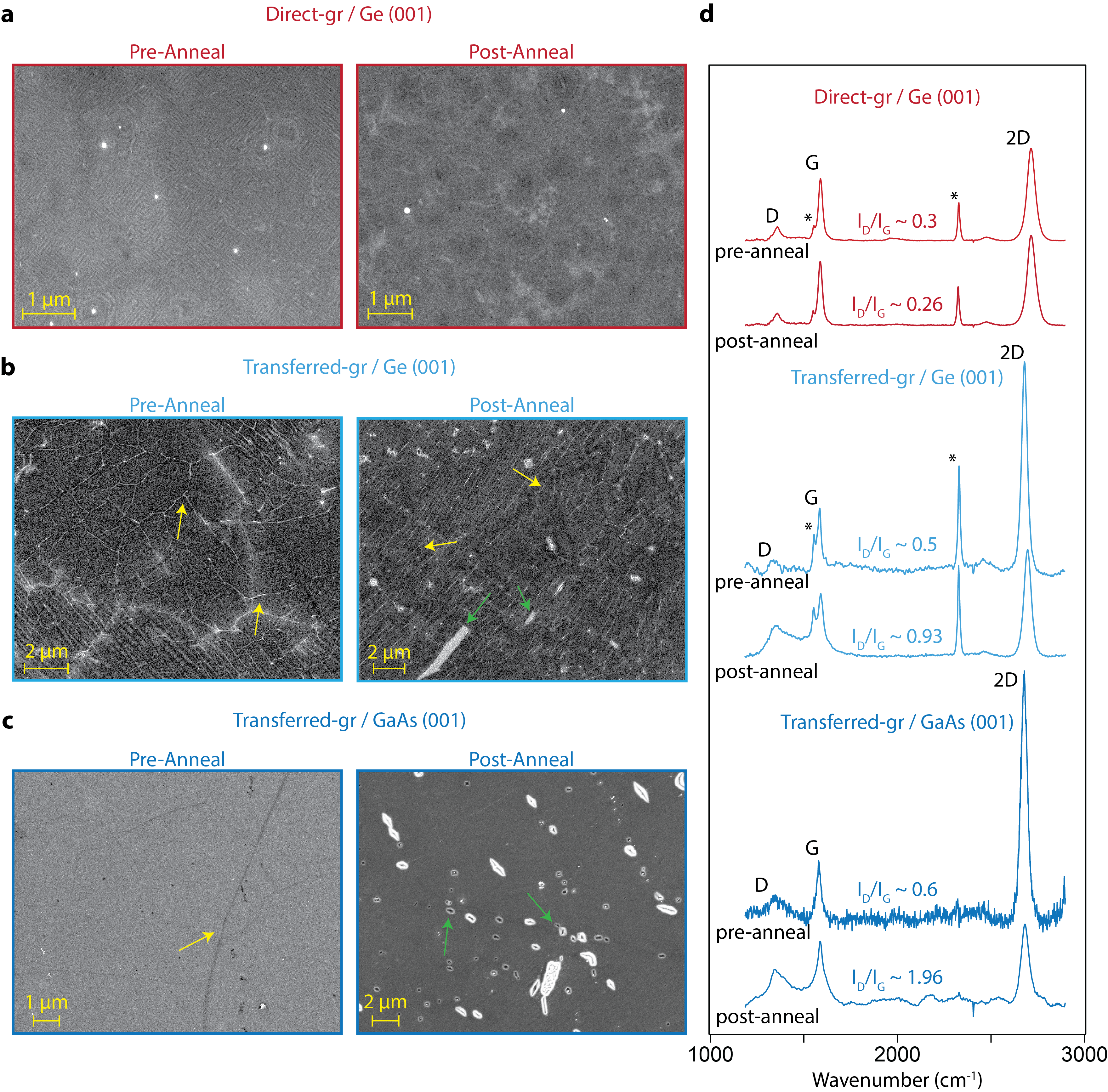}
    \caption{\textbf{Defects in direct-growth vs layer transferred graphene.} ``Pre-anneal'' refers to samples lightly outgassed at $300^\circ$C to remove adsorbates. ``Post-anneal'' refers to samples annealed at $600^\circ$C in ultrahigh vacuum to remove substrate native oxides.
    (a) SEM images for the direct-growth graphene (gr) on Ge (001) substrates before and after annealing.
    (b) SEM images for the layer-transferred graphene on Ge (001).
    (c) SEM images for the layer-transferred graphene on GaAs (001). Yellow arrows indicate wrinkles and grain boundaries, green arrows denote tears and holes in the graphene layer.
    (d) Raman spectra before and after annealing. The peaks marked by * are from the Ge substrate. For a description of the $D$ and $G$ peak fitting, see Supporting Information Fig. \ref{sm_ramanstruc}.}
    \label{graphene_char}
\end{figure}

\clearpage \newpage

Fig. \ref{graphene_char} illustrates the differences between the directly-grown graphene versus graphene that has been layer transferred onto Ge. We first examine the samples after light outgassing at $300^\circ$C to remove solvents and adsorbates, but before full annealing (labelled ``pre-anneal'', Fig. \ref{graphene_char}). For direct-growth graphene on Ge, the primary feature observed by scanning electron microscopy (SEM) is faceting of the Ge (001) surface (Fig. \ref{graphene_char}a, left), as has been observed previously \cite{mcelhinny2016graphene}. In contrast, the layer-transferred graphene on Ge displays a network of grain boundaries and wrinkles as labelled by the yellow arrows (Fig. \ref{graphene_char}b, left). The grain boundaries originate from graphene growth on polycrystalline Cu foils (Supporting Information Fig. \ref{sm_cufoilSEM}). Atomic force microscopy (AFM) measurements confirm the grain boundaries and wrinkles for transferred graphene (Fig. \ref{sm_afm}b, left). Neither layer transferred nor directly grown graphene show obvious signs of holes or tears prior to annealing in SEM or AFM measurements. Raman spectroscopy (Fig. \ref{graphene_char}d) shows the $I_{D}/I_{G}$ integrated intensity ratio for the direct and layer transferred graphene on Ge before anneal are similar (0.3 and 0.5, respectively), indicative of a similar defect density \cite{tuinstra1970raman, cancado2011grdefects}.  Layer-transferred graphene on GaAs (001) exhibits qualitatively similar surface morphology and Raman spectra as the graphene transferred onto Ge (001) (Fig. \ref{graphene_char}c, left). 

After annealing in ultrahigh vacuum at 600$^\circ$C, which is necessary to remove substrate native oxides prior to film deposition \cite{springthorpe1987measurement, oh2004thermal}, we observe stark contrast between the two different graphene preparations. For directly grown graphene on Ge, our Raman, SEM, and AFM measurements display negligible differences before and after annealing (Fig. \ref{graphene_char}a, right). AFM measurements reveal faint wrinkles in the graphene layer formed after annealing (Supporting Information Fig. \ref{sm_afm}), but no obvious formation of tears (Fig. \ref{graphene_char}a) and negligible changes in the Raman $I_D/I_G$ ratio (Fig. \ref{graphene_char}d, red curves). In contrast, for layer-transferred graphene our SEM and AFM measurements reveal holes and tears that appear to originate at grain boundaries (green arrows in Fig. \ref{graphene_char}b (right) and Supporting Information Fig. \ref{sm_afm}). This increase in defect density is further confirmed through Raman spectroscopy, where we observe a dramatic increase in the Raman $I_D/I_G$ ratio for transferred graphene after annealing (Fig. \ref{graphene_char}d, blue curves), indicating the creation of tears and other defects. We observe a similar increase in the tear density and Raman $I_D/I_G$ ratio for transferred graphene on GaAs after annealing (Fig. \ref{graphene_char}c).

We hypothesize that the anneal-induced tears are created by desorption of native oxides that are trapped between the substrate and layer transferred graphene. Previous in-situ XPS measurements demonstrate that layer transferred graphene on III-V substrates contains oxides trapped at the graphene/III-V interface \cite{manzo2021defect}. Desorption of these interfacial oxides creates pinholes in the graphene. We expect transferred graphene/Ge to behave similarly. In contrast, graphene that is directly grown on Ge does not have these interfacial contaminants, and thus does not show an increased defect density after annealing.

\begin{figure*}[ht]
    \includegraphics[width=0.9\linewidth]{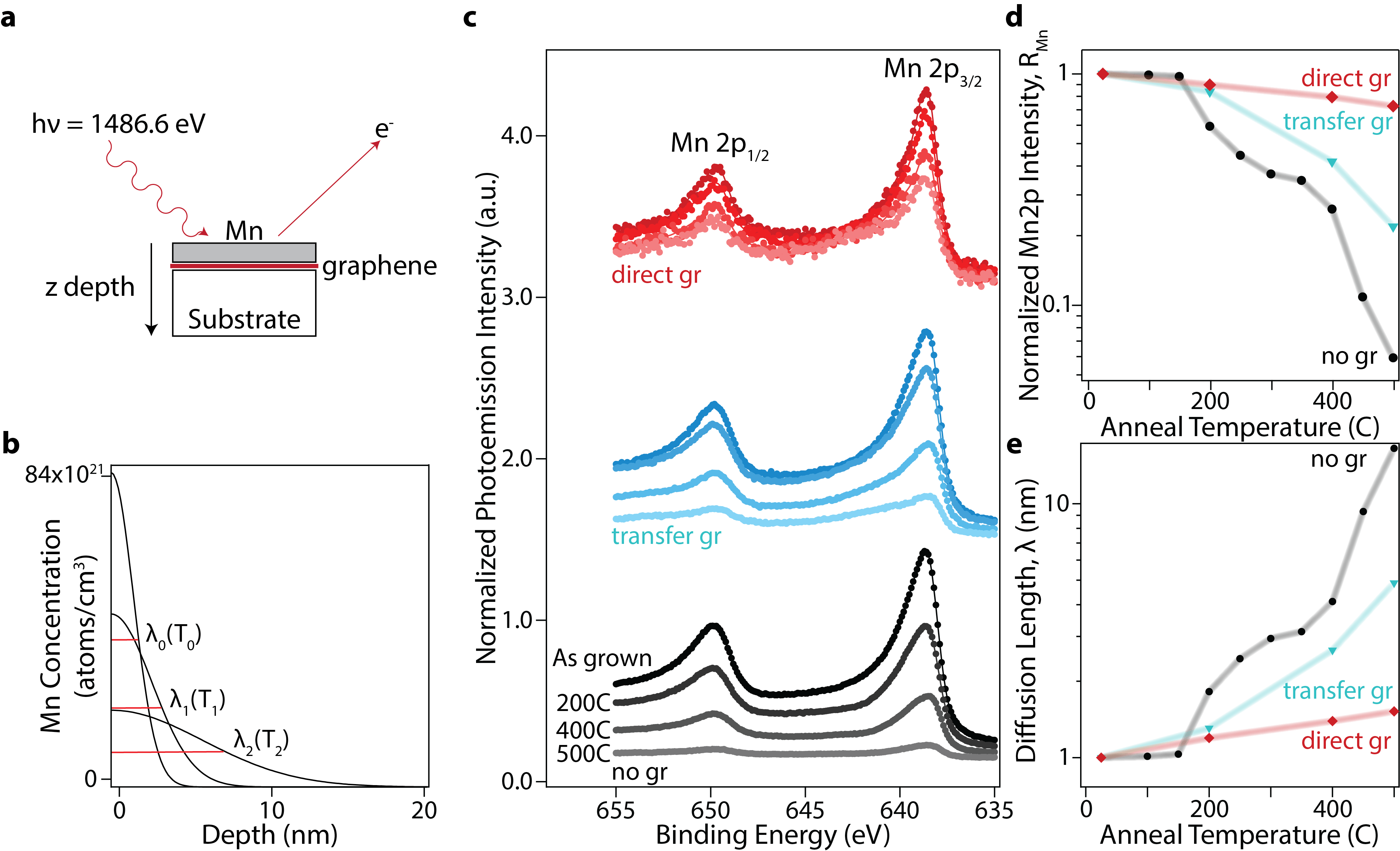}
	\caption{\textbf{In-situ photoemission measurements of Mn on graphene(gr)/Ge.}
	(a) Schematic layer structure and measurement geometry.
	(b) Half Gaussian composition profiles as a function of depth from the surface. We use this model to extract diffusion length $\lambda$.
	(c) Mn 2p core level spectra for Mn thin films grown directly on bare Ge (black), transferred graphene/Ge (blue), and direct-growth graphene on Ge (red) at increasing anneal temperatures, using an Al $K\alpha$ source ($h\nu = 1486.6$ eV). Samples were rapidly heated to the anneal temperature, held for 5 minutes, rapidly cooled, and measured at room temperature. Symbols are experimental data and curves are fits to a Doniach-Sunjic lineshape.
	(d) Normalized Mn intensity ratio $R_{Mn} = \frac{(I_{Mn2p}/I_{sub})_T}{(I_{Mn2p}/I_{sub})_0}$, tracking the relative change in Mn surface composition as a function of anneal temperature $T$. For substrate normalization we use $sub$ = Ge $3p$.
	(e) Calculated Mn diffusion lengths $\lambda$ at each anneal temperature.
	}
    \label{Ge_data}
\end{figure*}

\begin{figure*}[ht]
    \includegraphics[width=0.7\linewidth]{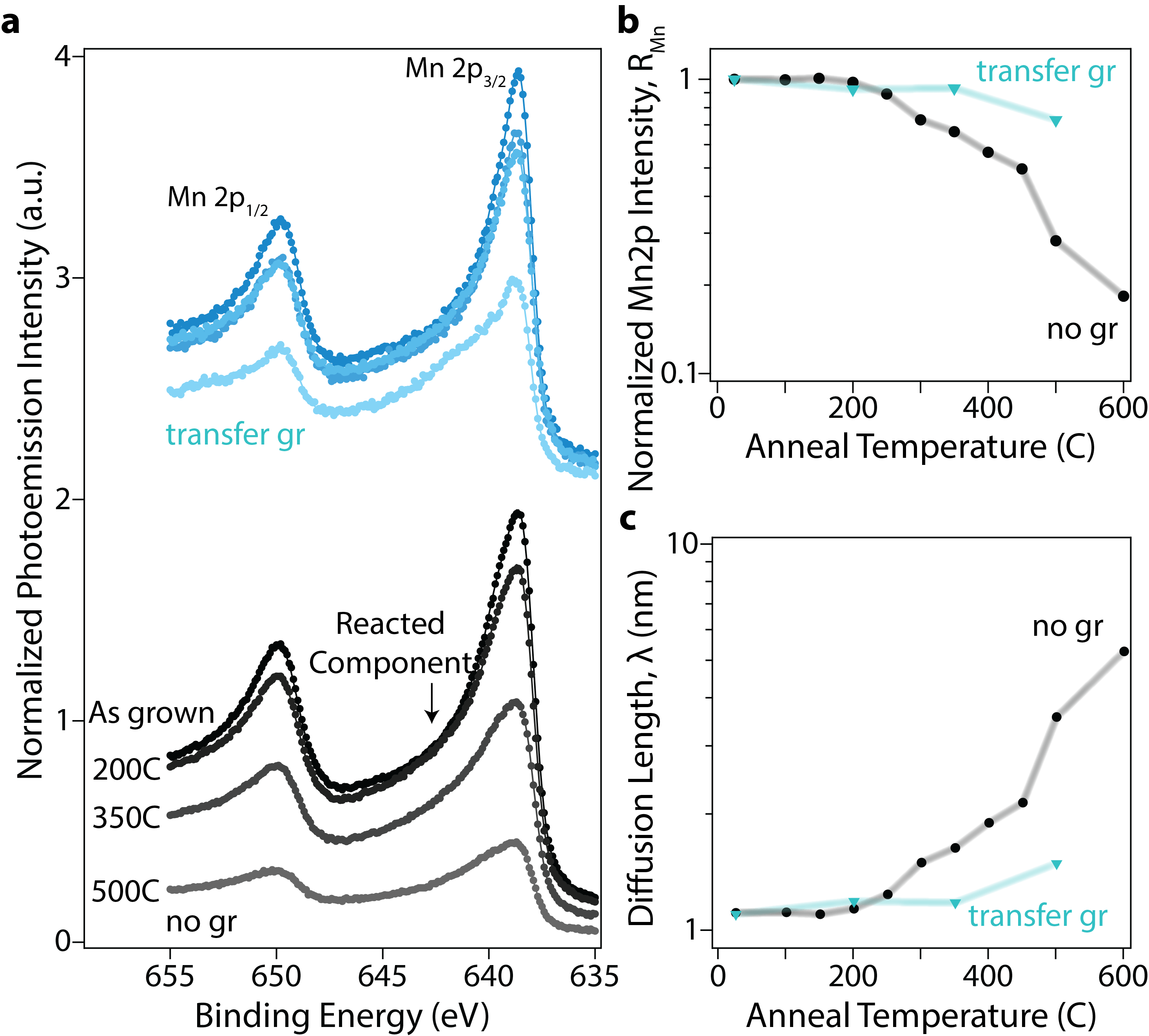}
    \caption{\textbf{In-situ photoemission measurements of Mn on graphene(gr)/GaAs.} (a) Core level spectra and fits for Mn/GaAs (black) and Mn/transferred-gr/GaAs (blue) at increasing anneal temperatures. Data points represent experimental photoemission results and the traces represent the fits. (b) Normalized Mn intensity $R_{Mn} = \frac{(I_{Mn2p}/I_{sub})_T}{(I_{Mn2p}/I_{sub})_0}$, tracking the relative change in Mn surface composition as a function of anneal temperature $T$. For substrate normalization we use $sub$ = Ga 3p. (c) Calculated diffusion lengths at each anneal temperature.
    }
    \label{GaAs_data}
\end{figure*}

We now measure the diffusion barrier performance for Mn thin films grown on the above graphene-terminated samples. After loading the graphene/Ge samples into vacuum and annealing at 600 C, Mn thin films ($1.64 \times 10^{16}$ atoms/cm$^2$, $\sim 24$ \AA) were grown by molecular beam epitaxy (MBE) at room temperature. After growth, samples were transferred via an ultrahigh vacuum manifold to a connected X-ray photoemission spectroscopy (XPS) chamber. In this way, we measure clean XPS spectra of samples that have not been exposed to air. Further details are described in Methods.

Figure \ref{Ge_data}c presents \textit{in-situ} photoemission spectra tracking the evolution of Mn surface composition for Mn films grown on Ge substrates, with and without a graphene barrier. Each sample was subjected to a series of anneal steps at increasing temperature up to 500 $^\circ$C. To eliminate variations in sample alignment, all spectra are normalized to the integrated Ge $3p$ core level intensity. For the Mn film grown directly on Ge (black curves), with increasing anneal temperature the intensity of the Mn $2p$ rapidly decreases, corresponding to a decrease in the relative surface Mn composition. For Mn on transferred graphene (blue curves) and on direct-growth graphene (red curves), the Mn $2p$ core level decay is suppressed. 

To compare the relative changes in surface Mn concentration, we fit the Mn $2p$ and Ge $3p$ core levels to Doniach-Sunjic \cite{doniach1970} and Voigt lineshapes, respectively, and plot the relative change in Mn / Ge intensity ratio, $R_{Mn} = \frac{(I_{Mn2p} / I_{sub})_T}{(I_{Mn2p} / I_{sub})_0}$ (Fig. \ref{Ge_data}d, $sub =$ Ge $3p$). Here the numerator with subscript $T$ refers to the Mn $2p$ / Ge $3p$ intensity ratio after anneal at temperature $T$ and the denominator with subscript $0$ refers to the Mn $2p$ / Ge $3p$ intensity ratio as-grown, before additional annealing. This normalization procedure cancels out the instrumental sensitivity, alignment, and photoemission cross section. 

We find that the directly-grown graphene performs best as a diffusion barrier, with $R_{Mn}$ decreasing to only 0.7 after an anneal up to 500$^{\circ}$C, compared to $R_{Mn}=0.05$ for the sample without graphene. Transferred graphene also suppresses interdiffusion compared to the bare substrate. For the no graphene sample there is a distinct onset of interdiffusion around 150$^\circ$C, at which point $R_{Mn}$ rapidly decreases. There is no distinct onset for the samples with graphene barriers. Together, these results demonstrate that both transferred and directly-grown graphene are effective solid state diffusion barriers for Mn in Ge, and that directly-grown graphene is most effective.

We next estimate the temperature-dependent diffusion length based on the relative changes in the Mn $2p$ / Ge $3p$ intensity ratio, $R_{Mn}$. For a very thin Mn film on a semi-infinite substrate, the solution to the diffusion equation is a half Gaussian composition profile (Fig. \ref{Ge_data}b)
\begin{equation}
    c_{Mn}(z,\lambda) = \frac{n_{total}}{\sqrt{\pi} \lambda} exp \left(\frac{-z^2}{4 \lambda^2} \right)
    \label{Cx}
\end{equation}
where $\lambda$ is the diffusion length and $n_{total} = 1.64 \times 10^{16}$ atoms/cm$^{2}$ is the total areal density of Mn deposited. We define the as-deposited initial condition such that the Mn concentration at the surface ($z=0$) is equal to the density of pure Mn, i.e. $c_{Mn}(0,\lambda_0) = c_0= 8.2\times 10^{22}$ atoms/cm$^{3}$. This yields an initial diffusion length of $\lambda_0 = 11$ \AA. For simplicity, we let the Ge composition profile be $c_{Ge}(z) = c_{0} - c_{Mn}(z)$, assuming that pure Mn, pure Ge, and the diffused Mn-Ge have approximately the same atomic densities and layer spacings.

From these composition profiles we calculate the expected intensity ratios for Mn 2p and Ge 3p, as a function of the diffusion length. The intensity of core level $x$ is given by
\begin{equation}
    I_{x} = f_{x}\sigma_{x}\sum_{i=0}^{\infty}n_{x,i} exp \left( \frac{-z_i }{\lambda_{x,imfp} cos\phi} \right)
    \label{xps_int}
\end{equation}
where $n_{x,i}$ is the two-dimensional density of atomic layer $i$, $z_i$ is the depth from the surface, $\lambda_{x,imfp}$ is the  inelastic mean free path (20 \AA\ for Mn $2p$, 26 \AA\ for Ge $3p$, 22.1 \AA\ for Ga $3p$ \cite{tanuma2003calculation}), $\phi=0^\circ$ is the emission angle from surface normal, $f_x$ is an instrument sensitivity factor, and $\sigma_x$ is the photoemission cross section. We rewrite the composition profiles in terms of atomic density, $n_{x,i} = d * c_{x}(z,t)$, where $d$ is the atomic layer spacing, and calculate the photoemission intensities $I_{Mn}$ and $I_{Ge}$, subject to the initial condition $\lambda_0 = 11$ \AA. We then compute the normalized Mn intensity ratio $R_{Mn}=\frac{ (I_{Mn} / I_{Ge})_T}{(I_{Mn} / I_{Ge})_0}$, such that $f_x$ and $\sigma_x$ are cancelled out. In this way, the experimentally measured $R_{Mn}$ is uniquely defined by the diffusion length $\lambda$ after each sequential anneal.

Fig. \ref{Ge_data}(e) shows the estimated diffusion length, obtained by fitting the experimentally measured $R_{Mn}$ to the calculated intensity ratios from Eqs 1 and 2. We find that compared to the bare Ge substrate, directly-grown graphene on Ge suppresses the diffusion length by a factor of just under 11 and layer transferred graphene suppresses the diffusion length by a factor 3.5, at a temperature of 500 $^\circ$C.

In-situ photoemission measurements for Mn on GaAs substrates reveal similar behavior as the samples on Ge (Fig. \ref{GaAs_data}). For the Mn directly on GaAs (Fig. \ref{GaAs_data}a, black curves), in the as-grown spectrum we observe a secondary component at higher binding energy, indicative of reactions or secondary phases between Mn and GaAs. This reacted component grows with increased annealing. We attribute these secondary components to the formation of Mn$_{2}$As and MnGa phases, as has been observed previously at Mn/GaAs interfaces without graphene \cite{hilton2004}. Comparison of the normalized Mn 2p intensity $R_{Mn}$ and the estimated diffusion length reveal that like Ge, transferred graphene suppresses interdiffusion into GaAs.

\begin{figure*}[ht]
    \includegraphics[width=0.7\linewidth]{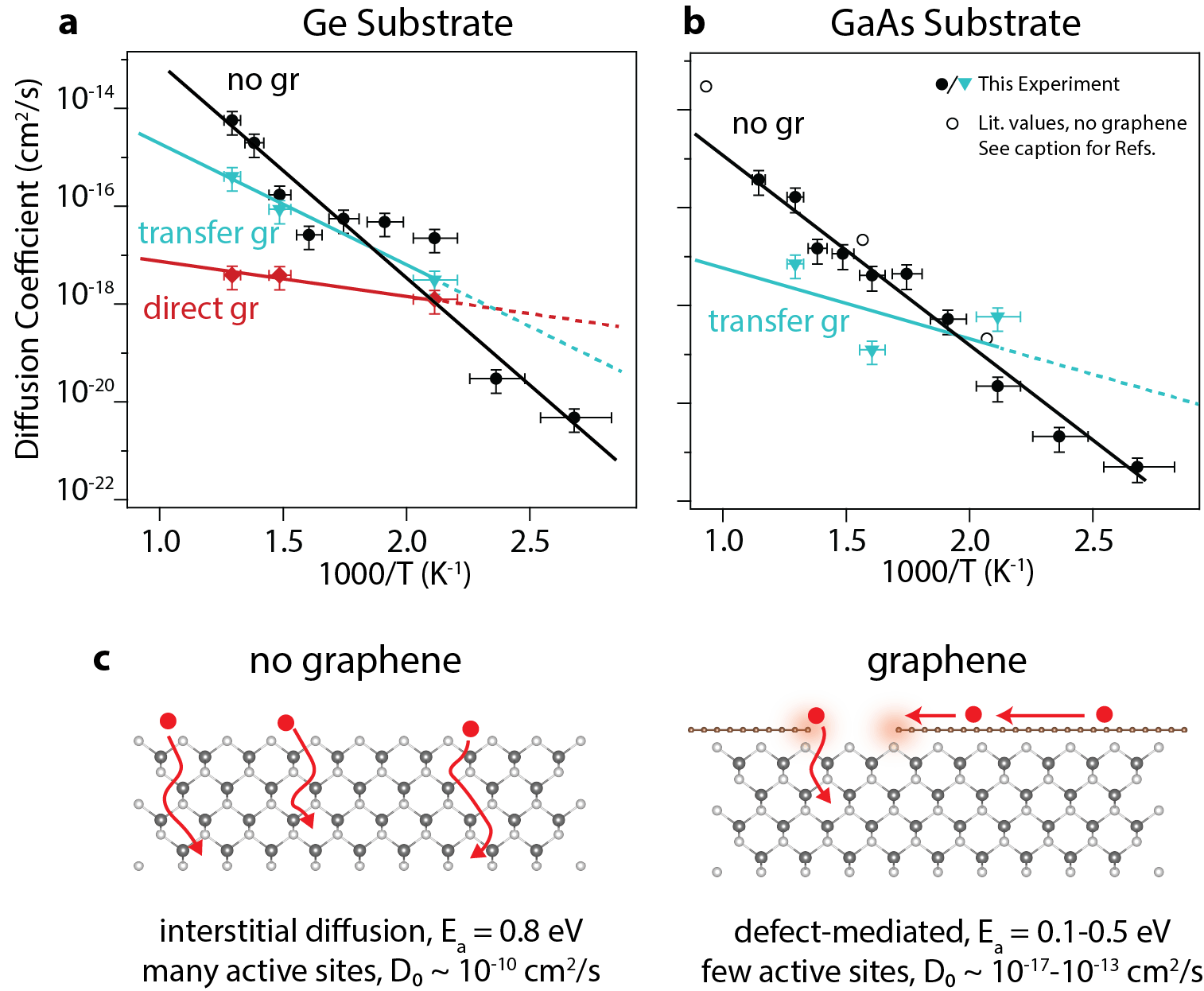}
	\caption{\textbf{Calculated diffusion coefficients for Mn on Ge and GaAs, with and without graphene (gr).} (a) Diffusivity vs inverse temperature for Mn/Ge (black), Mn/layer transferred graphene/Ge (blue), and Mn/direct-growth graphene/Ge (red). Symbols are experimental data, the line is an Arrhenius fit. 
	(b) Diffusivity vs inverse temperature for Mn/GaAs (black) and Mn/layer transferred graphene/GaAs (blue). The open symbols represent previously reported diffusivity for Mn on GaAs (no graphene) from Refs. \cite{seltzer1965,adell2011,vikhrova2012}.
	(c) Schematic diffusion processes.\\
	}
    \label{DvT}
\end{figure*}

We now use the diffusion length to estimate the diffusivity via $\lambda^2 = Dt$. Since each sample is subjected to a series of anneals at increasing temperature, we approximate the cumulative diffusion length squared after $N$ anneal steps as a sum of the previous diffusion steps
\begin{equation}
    \lambda_{total,N}^2 = \sum_{i=1}^{N}{D(T_i) \Delta t_i},
    \label{Eq1_lambdasum}
\end{equation}
where $\Delta t_i$ is the time increment of anneal step $i$ and $D(T_i)$ is the diffusivity from the anneal at temperature $T_i$. Since each anneal is for a constant amount of time ($\Delta t_i = $ 5 minutes), we rewrite Equation \ref{Eq1_lambdasum} to solve for $D(T_i)$ contribution of an individual anneal step.
\begin{equation}
\begin{aligned}
    D(T_{i=N}) &= \frac{\lambda^2_{total,N}}{\Delta t} - \sum_{m=1}^{N-1}D(T_m) \\
\end{aligned}
\end{equation}

We plot the diffusivity $D(T_i)$ as a function of temperature in Fig. \ref{DvT}. From an Arrhenius fit, $D(T) = D_0 exp(-E_a/k_B T)$, we extract activation energies $E_a$ and diffusivity prefactors $D_0$ as listed in Table I. We find that at 500 $^\circ$C, the diffusivity for Mn on directly-grown graphene / Ge (red curve) is a factor of 1000 lower than the diffusivity on bare Ge (black curve). Mn on transferred graphene / Ge is only a factor of 10 lower (blue curve). Similar behavior is observed for transferred graphene/GaAs compared to bare GaAs. To benchmark the accuracy of our method, we find that our experimentally determined diffusivities for Mn on bare GaAs (Fig. \ref{DvT}(b), filled black circles) are in good agreement with previous experiments (open black circles) \cite{seltzer1965,adell2011,vikhrova2012}. 

\begin{table}[htbp]
\begin{tabular}{l|c|cc|c} 
                    & $I_{D}/I_{G}$ & $D_0$ (cm$^2$/s)     & $E_a$ (eV) & $D(500^\circ$C) (cm$^2$/s) \\
                    & post-anneal & & & \\
\hline
Ge                  & -- & $6.0\times 10^{-10}$ & 0.81       & $5\times10^{-15}$ \\
transfer gr/Ge      & 0.93 & $6.0\times 10^{-13}$ & 0.50       & $4\times10^{-16}$ \\
direct-growth gr/Ge     & 0.26 & $3.0\times 10^{-17}$ & 0.13       & $4\times10^{-18}$ \\ 
\hline
GaAs                & -- & $1.3\times 10^{-11}$ & 0.79       & $4\times10^{-16}$ \\
transfer gr/GaAs    & 1.96 & $5.2\times 10^{-17}$ & 0.21       & $7\times10^{-18}$ 
\end{tabular}
\caption{Diffusivity refactors $D_0$, activation energies $E_a$, and diffusivity at 500 $^\circ$C, for Mn diffusion into Ge and GaAs, extracted from in-situ XPS. The graphene $I_{D}/I_{G}$ integrated intensity ratio is from Raman Spectroscopy before Mn film growth (Fig. \ref{graphene_char})}.  
\label{tab_diff}
\end{table}


We now discuss the mechanisms for Mn diffusion based on the experimentally determined $E_a$ and $D_0$. Our experimentally determined activation energy of 0.8 eV for Mn diffusing into bare GaAs and Ge agrees well with previous experiments and DFT calculations for Mn interstitial diffusion into GaAs ($0.7-0.8$ eV \cite{edmonds2004}). Substitutional diffusion is expected to exhibit a higher activation energy of 2-3 eV \cite{raebiger2006}. 

In comparison, $E_a$ for Mn diffusion into graphene-terminated substrates is considerably lower: 0.1 eV for directly-grown graphene on Ge, and 0.2-0.5 eV for transferred graphene on GaAs and Ge. We immediately rule out transverse diffusion through pristine graphene, since diffusion through graphene in the absence of defects is expected to have a very large activation energy barrier of $E_a \sim 12$ eV \cite{leenaerts2008}. This is because the effective pore size of graphene is small ($\sim 0.7$ \AA\ radius of a carbon ring, less than half the radius of a Mn atom). Therefore, we expect that Mn diffusion through graphene occurs primarily through defect-mediated processes.

The range of measured activation energies suggests several mechanisms for Mn diffusion through graphene defects may be possible. At high temperatures ($T>200^\circ$ C, or $1000/T < 2.1$ K$^{-1}$), we speculate that our extracted $E_a$ is lower on graphene-terminated substrates than on bare substrates due to reactions being catalyzed at graphene defects, since graphene defects are known to be chemically active \cite{roy2013improving, yoon2014cuox}. At this point we do not have sufficient data to determine the mechanisms of diffusion lower temperatures, $T<200^\circ$ C (or $1000/T > 2.1$ K$^{-1}$, Fig. \ref{DvT} dotted lines). The range of measured activation energies may also reflect a convolution of surface and bulk diffusion, in which Mn adatoms first diffuse laterally to reactive pinhole sites before then diffusing into the substrate (Fig. \ref{DvT}c). First-principles calculations suggest that the activation energy for lateral diffusion of Mn on graphene is 60 meV \cite{liu2016growth} and Cu on graphene is 10-20 meV \cite{han2019energetics}, similar to the activation energies extracted from our photoemission measurements (Table \ref{tab_diff}). More detailed experiments with lateral and depth resolution, at a broader range of temperatures, are needed to fully understand the detailed diffusion mechanisms through graphene.

While a low activation energy barrier favors a large diffusivity, the origin of the dramatic decrease of diffusivity for graphene-terminated samples is the reduction of the prefactor $D_0$. We interpret $D_0$ to scale as an attempt rate for diffusion. For Mn on bare GaAs and Ge substrates, the attempt rates of $10^{-11}-10^{-10}$ cm$^2$/s are high because there are many available sites for interstitial diffusion into the substrate. On the other hand, attempt rates are lower on the graphene-terminated substrates because Mn diffusion through graphene requires a defect in the graphene. This rate is higher for layer transferred graphene ($10^{-17}-10^{-13}$ cm$^2$/s) than for directly-grown graphene ($10^{-17}$ cm$^2$/s), since the layer-transferred graphene has a higher defect density as quantified by our Raman, SEM, and AFM measurements (Fig. \ref{graphene_char} and Supporting Information Fig. \ref{sm_afm}). Note that film/substrate interactions \cite{obraztsov2008raman} and doping \cite{das2008monitoring} can modify Raman intensity ratios \cite{ferralis2010probing}, therefor it is difficult to make direct comparisons of gr/Ge and gr/GaAs defect densities based on the $I_D/I_G$ ratio (Fig. \ref{graphene_char}d). Further experiments are required to correlate specific defect types (e.g. tears, wrinkles, pinholes, vacancies) with specific values of the diffusivity.


\section{Conclusions}

Our results show that layer-transferred and directly-grown graphene suppress Mn diffusion into semiconductor substrates. The directly-grown graphene displays superior diffusion barrier performance due to a lower defect density, as quantified by Raman spectroscopy and SEM after annealing but before Mn film growth. Translating these performance enhancements to functional substrates at wafer scale will require direct graphene synthesis approaches on the substrate of interest. For example, CVD approaches using small molecules \cite{li2011low}, activated precursors \cite{lee2010laser}, or plasma-enhanced CVD processes \cite{vishwakarma2019pcvd, mischke2020ganpcvd, li2021pcvd} show promise for reducing the graphene synthesis temperature, and may be a route for graphene growth directly on GaAs and other technologically important compound semiconductors. 

\textbf{Supporting Information:} SEM images of graphene grown on Cu foils; AFM measurements of directly grown graphene on Ge, transferred graphene on Ge, and transferred graphene on GaAs; and Raman fitting of the graphene $D$ and amorphous carbon peaks.

\section{Acknowledgment}

Mn film growth, graphene transfer, and photoemission experiments by PJS, SM, and JKK were primarily supported by the CAREER program National Science Foundation (DMR-1752797). In-situ photoemission experiments and analysis by PJS and JKK was partially supported by the NSF Division of Materials Research through the University of Wisconsin Materials Research Science and Engineering Center (Grant No. DMR-1720415). Graphene synthesis and characterization by VS, KS, and MSA are supported by the U.S. Department of Energy, Office of Science, Basic Energy Sciences, under award no. DE-SC0016007. We gratefully acknowledge the use of Raman and electron microscopy facilities supported by the NSF through the University of Wisconsin Materials Research Science and Engineering Center under Grant No. DMR-1720415.


\section{References}
\providecommand{\latin}[1]{#1}
\makeatletter
\providecommand{\doi}
  {\begingroup\let\do\@makeother\dospecials
  \catcode`\{=1 \catcode`\}=2 \doi@aux}
\providecommand{\doi@aux}[1]{\endgroup\texttt{#1}}
\makeatother
\providecommand*\mcitethebibliography{\thebibliography}
\csname @ifundefined\endcsname{endmcitethebibliography}
  {\let\endmcitethebibliography\endthebibliography}{}

\end{document}


\renewcommand{\thetable}{S-\arabic{table}}   
\renewcommand{\thefigure}{S-\arabic{figure}}

\begin{figure}[ht]
    \includegraphics[width=0.5\linewidth]{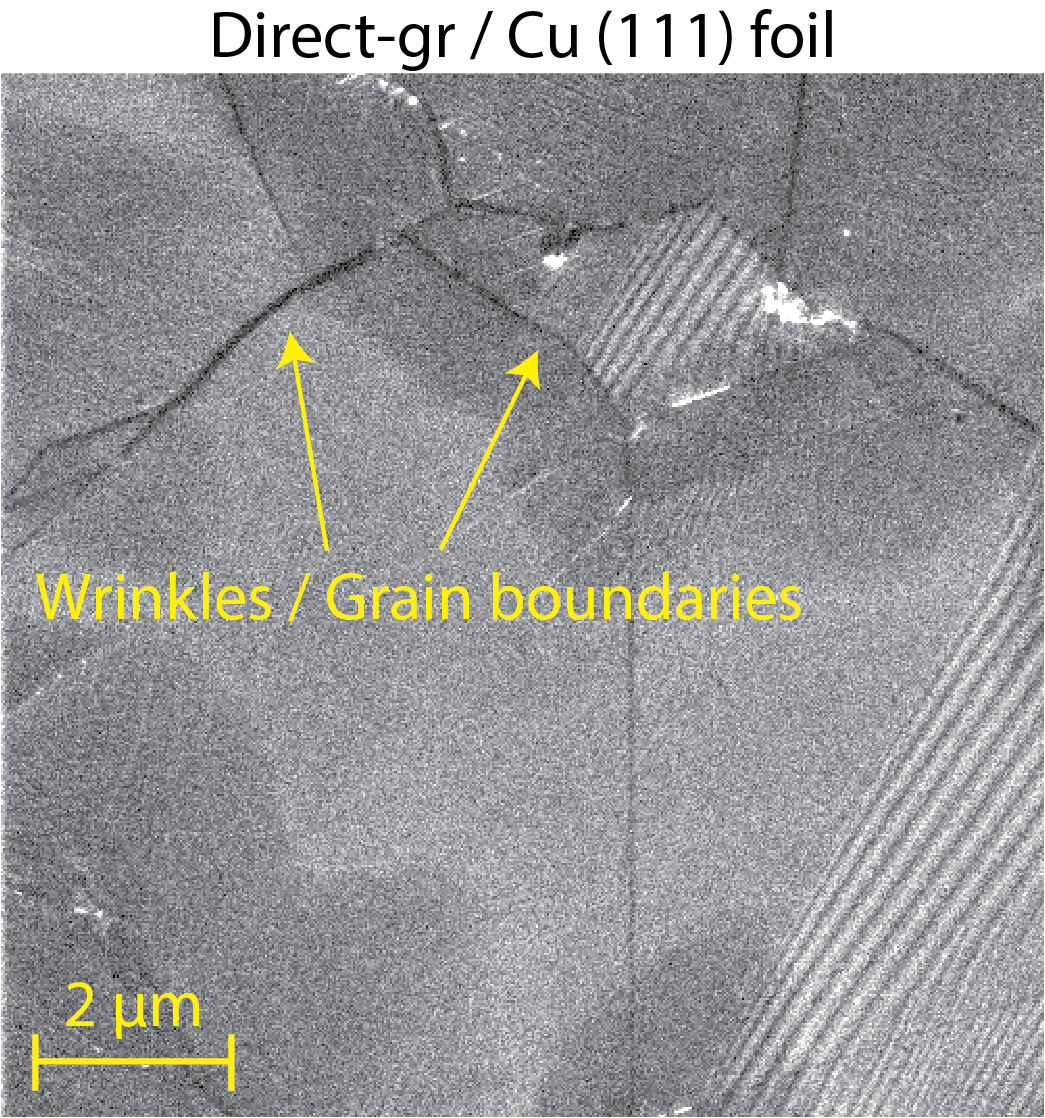}
    \caption{\textbf{SEM of CVD graphene grown on Cu foil.}
    Plan-view scanning electron microscopy (SEM) image of the graphene as-grown on copper foil via standard chemical vapor deposition (CVD) growth techniques. The main structure observed in the graphene is a few large-scale wrinkles and grain boundaries. The well-ordered parallel line structure observed is a result of the underlying structure of the Cu foil.}
    \label{sm_cufoilSEM}
\end{figure}

\begin{figure}[ht]
    \includegraphics[width=0.95\linewidth]{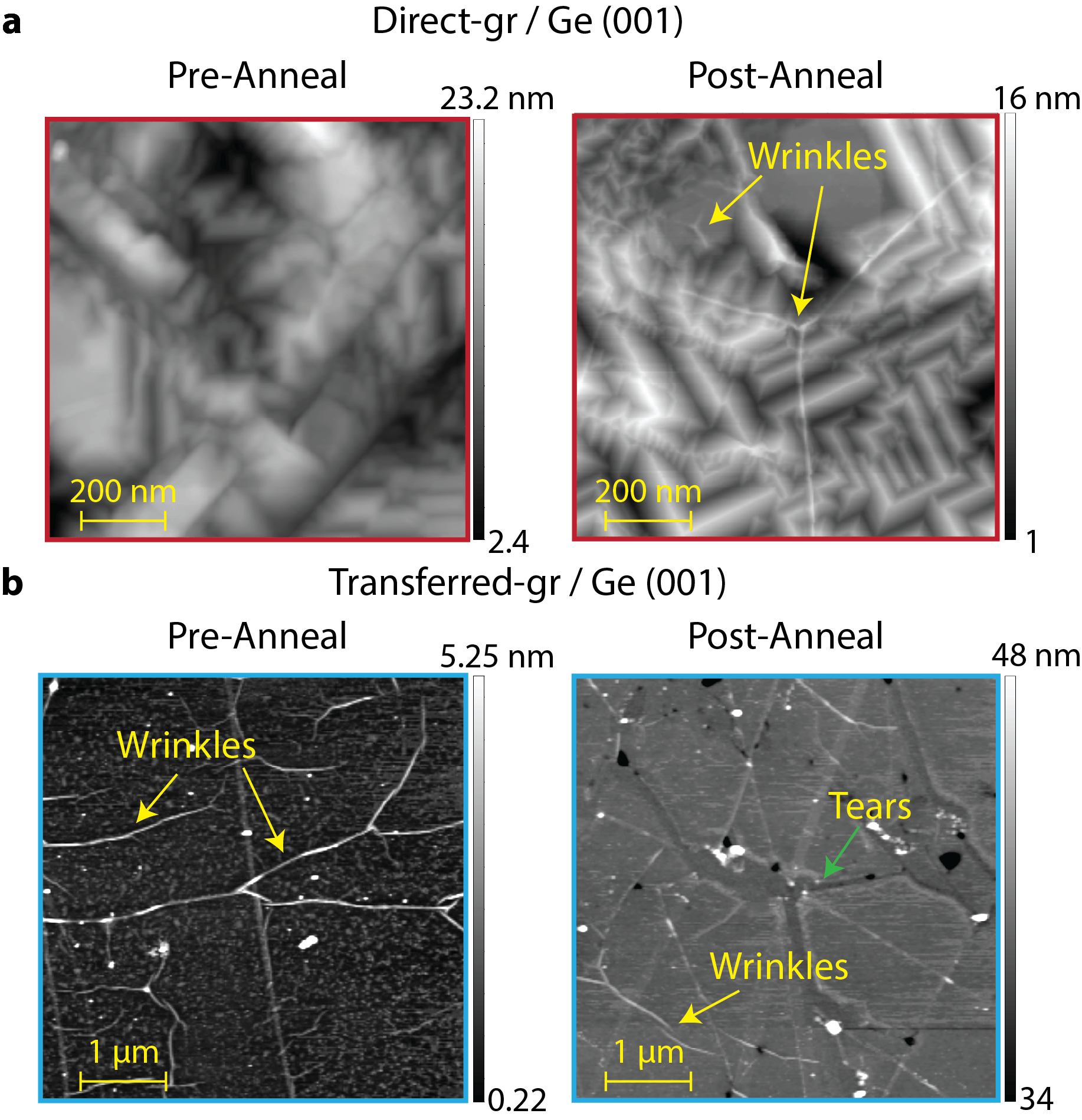}
    \caption{\textbf{AFM characterization of gr/Ge (001) surfaces.}
    (a) Atomic force microscopy (AFM) images of the graphene directly grown on Ge (001) substrates before (left) and after (right) annealing. Prior to annealing, the morphology is dominated by facets from the underlying Ge (001) \cite{mcelhinny2016graphene}. Features from the graphene are less obvious. After annealing at 600$^\circ$C, the graphene develops large-scale wrinkles; however, no noticeable tears or holes have formed.
    (b) AFM images of transferred graphene on Ge (001) substrates before (left) and after (right) annealing. Prior to annealing the major defect structure is the wrinkles and grain boundaries observed originally via SEM prior to the transfer process (Fig \ref{sm_cufoilSEM}). The underlying terrace structure of the Cu foil also produces a very small ripple-like wrinkle in the graphene at regular intervals. After annealing, the graphene appears to rip apart at some of the grain boundaries, forming a network of tears in the graphene.
    }
    \label{sm_afm}
\end{figure}


\begin{figure}[ht]
    \includegraphics[width=0.8\linewidth]{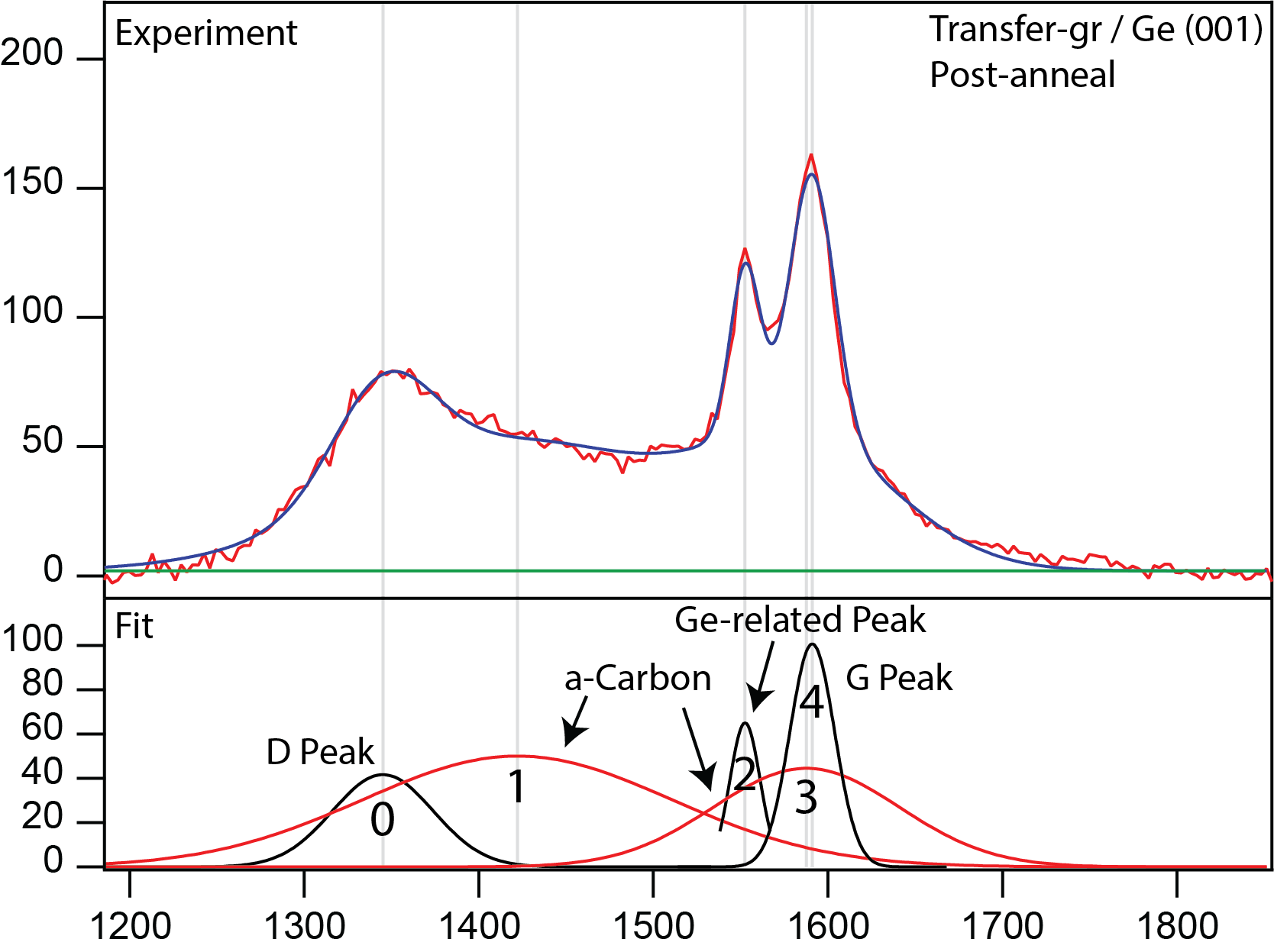}
    \caption{\textbf{Deconvolution near the Raman $D$ peak for transferred graphene.}
    For transferred graphene after annealing, we observe broad peaks (labelled 1 and 3) in the vicinity of the graphene $D$ peak. We attribute these broad peaks to amorphous carbonaceous species from polymer residuals, consistent with previous studies \cite{hong2013origin}. To extract the $D$ component (labelled 0) and quantify an integrated $I_D / I_G$ ratio, we fit the data using multiple Gaussian lineshapes.\\
    }
    \label{sm_ramanstruc}
\end{figure}

\clearpage
\newpage
\bibliography{bibliography}